\documentclass[prl,aps,twocolumn,floatfix,showpacs]{revtex4}
\usepackage{graphicx,graphics,psfrag,amsmath,calc}
\usepackage{epsfig}
\usepackage{color}
\begin{document}

\title{Quantum phases of interacting three-component fermions \\
under the influence of spin-orbit coupling and Zeeman fields}

\author{
Doga Murat Kurkcuoglu and C. A. R. S{\'a} de Melo
}

\affiliation{
School of Physics, Georgia Institute of Technology, Atlanta, 
Georgia 30332, USA
}

\date{June 8, 2016}
\pacs{03.75.Ss, 67.85.Lm, 67.85.-d}

\begin{abstract}
We describe the quantum phases of interacting three component fermions
in the presence of spin-orbit coupling, as well as linear and quadratic 
Zeeman fields. We classify the emerging superfluid phases 
in terms of the {\it loci} of zeros of their quasi-particle 
excitation spectrum in momentum space, and we identify several 
Lifshitz-type topological transitions. In the particular case 
of vanishing quadratic Zeeman field, a quintuple point
exists where four gapless superfluid phases with surface and line nodes
converge into a fully gapped superfluid phase. Lastly, we also show that 
the simultaneous presence of spin-orbit and Zeeman fields transforms 
a momentum-independent scalar order parameter into an explicitly 
momentum-dependent tensor in the generalized helicity basis.
\end{abstract}
\maketitle

%
%

The effects of spin-orbit coupling are ubiquitous in the physical
world ranging from planetary motion in the macroscopic scale to the
hyperfine structure of atoms in the microscopic universe. 
At the quantum level, spin-orbit coupling is a relativistic 
manifestation associated with the motion of electrons around 
the atomic nucleus~\cite{landau-1980}. In electronic solids, 
this has lead to non-trivial effects such as the emergence of
topological insulators and superconductors~\cite{hasan-2010,zhang-2011a}, 
as well as non-centrosymmetric superconductivity~\cite{sigrist-2009}. 
However, neither spin-orbit coupling nor interactions can be changed 
substantially for a given condensed matter system.

Recently, artificial spin-orbit coupling with {\it adjustable} 
strength were created in neutral ultra-cold atoms. 
This was first achieved in the context of bosonic atoms, 
such as $^{87}$Rb, by using a set of Raman beams to
transfer momentum to the center of mass of atoms depending on 
their internal hyperfine state~\cite{spielman-2011}. 
The same technique was also used for fermionic atoms, such as $^{40}$K, 
which have the added advantage of the {\it tunability} of  
interactions via Fano-Feshbach 
resonances~\cite{spielman-2013,jiang-2014,chapman-2011}. 
For $^{87}$Rb, the interactions can not be adjusted, 
but it has been possible to study experimentally~\cite{spielman-2011} 
and theoretically~\cite{wu-2011,ho-2011,stringari-2012} the emergence
of superfluid phases in the presence of simultaneous artificial spin-orbit
coupling and Zeeman fields for systems of two hyperfine states.
For $^{40}$K, heating from the Raman beams made it difficult to cool down 
these fermions to sufficiently low 
temperatures~\cite{spielman-2013,jiang-2014}. This has precluded the 
probing of superfluid phases that have been predicted 
theoretically~\cite{shenoy-2011,zhang-2011b,zhai-2011,pu-2011,han-2012,seo-2012}
for two hyperfine states, when the interactions are changed. 
Although, presently, finite temperature theories for spin-orbit coupled 
fermions with two hyperfine states have been developed only for 
two-dimensional systems~\cite{devreese-2014}, 
it has not yet been possible to reduce the temperature of the atomic gas 
and reach the superfluid regime in the presence of Raman beams.
This has been true even for the intermediate or strongly attractive regimes, 
where the critical temperature for superfluidity 
is a fraction of the Fermi energy.

Currently, a new experimental method is being developed
where artificial spin-orbit coupling and Zeeman fields can be 
created by a specially designed chip. In this case, spatially modulated
radio-frequency fields can be created and can couple directly 
to atoms in a nearby cloud~\cite{goldman-2010}. 
This procedure can avoid the deleterious thermal 
effects caused by Raman beams, and may allow for the exploration 
of superfluid phases of fermionic systems with two or more hyperfine states.
In anticipation of the use of this technique, 
we propose the exploration of fermionic systems with three 
internal (or hyperfine) states in the presence of spin-orbit 
and Zeeman fields, where novel superfluid phases emerge 
as discussed below.

%
%

To describe interacting three-component fermions under the influence of 
spin-orbit and Zeeman fields, we start with the most general 
independent-particle Hamiltonian that could result from  
a suitably designed radio-frequency chip or Raman beams 
in the rotating frame~\cite{kurkcuoglu-2015}
\begin{eqnarray}
\label{eqn:independent-particle-hamiltonian-matrix}
{\bf H}_0({\bf k})
= 
\left(
\begin{array}{ccc}
\varepsilon_1({\bf k}) 	& \Omega_{12}	         &   \Omega_{13}	  \\
\Omega_{12}^* 		& \varepsilon_2({\bf k}) &   \Omega_{23}	  \\
\Omega_{13}^* 		& \Omega_{23}^*	         & \varepsilon_3({\bf k}) \\
\end{array} 
\right),
\end{eqnarray}
where 
$
\varepsilon_s({\bf k}) = ({\bf k}-{\bf k}_s)^2/(2m) + \eta_s
$
represents the energy of internal state 
$
s = 
\{
1,2,3
\}
$
after net momentum transfer
${\bf k}_s$, and $\eta_s$ is 
a reference energy of the atom at internal state $s$. 
The matrix elements $\Omega_{s s^{\prime}}$ represent Rabi frequencies
between atomic states $s$ and $s^{\prime}$. 

In this manuscript, instead of pursuing the most general theoretical case
shown in Eq.~(\ref{eqn:independent-particle-hamiltonian-matrix}), 
we consider a simpler experimental setup, where the Rabi frequency 
$\Omega_{13} = 0$, indicating that there is no coupling between states
$1$ and $3$. In addition, we consider that the Rabi frequencies associated
with the transitions from states $1$ to $2$ and from $2$ to $3$ to real and 
equal, that is, 
$\Omega_{12} = \Omega_{12}^* = \Omega_{23} = \Omega_{23}^* = \Omega$.
Furthermore, we choose a symmetric situation, where momentum 
transfers occur only to state 1 and 3, such that 
${\bf k}_1 = k_T {\hat {\bf x}}$,
${\bf k}_2 = 0$,
and ${\bf k}_3 = -k_T {\hat {\bf x}}$,
where $k_T$ is the magnitude of the momentum transferred 
to the atom by the photons. 
Lastly, we can define an energy reference via the sum 
$
\sum_s \eta_s = \eta,
$
leading to internal energies 
$
\eta_1 = -\delta,
$
$
\eta_2 = \eta
$ 
and
$
\eta_3 = +\delta,
$
where $\delta$ represents the detuning.

A simple rearrangment of the chip-atom or light-atom 
interaction Hamiltonian matrix allows us to write it 
in a more compact and transparent notation as
\begin{equation}
\label{eqn:spin-one-hamiltonian}
{\bf H}_0({\bf k}) 
= 
\varepsilon ({\bf k}){\bf 1} - 
h_x({\bf k}){\bf J}_x - 
h_z({\bf k}){\bf J}_z + 
b_z {\bf J}_z^2 ,
\end{equation}
where ${\bf J}_{\ell}$, with $\ell = \{ x, y, x \}$, 
are spin-one angular momentum matrices. 
Here, 
$
\varepsilon ({\bf k}) = {\bf k}^2/(2m) + \eta
$ 
is a reference kinetic energy which is the same 
for all internal states,
$
h_x({\bf k}) = -\sqrt{2} \Omega
$ 
is the {\it spin-flip} (Rabi) field, and 
$
h_z({\bf k}) = 2k_T k_x /(2m) + \delta
$
is a momentum dependent Zeeman field along $z$-axis, 
which is transverse to the momentum transfer direction ($x$-axis), 
and
$
b_z = k_T^2/(2m) - \eta
$
is the quadratic Zeeman term. Notice that $h_z ({\bf k})$ contains
the spin-orbit coupling term $2k_T k_x /(2m)$ as well as the detuning
$\delta$. Very recently, a similar Hamiltonian was created 
experimentally for spin-one bosonic $^{87}$Rb 
atoms~\cite{spielman-2015} and magnetic phases of this system were 
investigated.

The chip-atom (or light-atom) interaction Hamiltonian 
can be written in second-quantized notation as 
\begin{equation}
\label{eqn:independent-particle-hamiltonian-operator}
{H}_{{\rm CA}}
= 
\sum_{\bf k}
{\bf \Psi}^{\dagger}_{\bf k}
{\bf H}_0({\bf k})
{\bf \Psi}_{\bf k}
\end{equation}
where the spinor operator is 
$
{\bf \Psi}^{\dagger}_{\bf k} 
= 
(
\psi_1^{\dagger}({\bf k}),
\psi_2^{\dagger}({\bf k}),
\psi_3^{\dagger}({\bf k})
)
$,
with
$
\psi_s^{\dagger}({\bf k})
$
creating a fermion with momentum 
${\bf k} - {\bf k}_s$ in internal state $s$. 
The Hamiltonian $H_{\rm CA}$
can be diagonalized via the rotation 
$
{\bf \Phi} ({\bf k}) 
= 
{\bf U}({\bf k}) 
{\bf \Psi} ({\bf k}),
$
which connects the three-component spinor ${\bf \Psi} ({\bf k})$ 
in the original spin basis to the three-component spinor 
${\bf \Phi} ({\bf k})$ 
representing the basis of eigenstates. The matrix ${\bf U}({\bf k})$ 
is unitary and satisfies the relation 
${\bf U}^\dagger ({\bf k}) {\bf U} ({\bf k}) = {\bf 1}$. 
The diagonalized Hamiltonian matrix is
$
{\bf H}_D({\bf k}) 
= 
{\bf U}({\bf k})
{\bf H}_{0}({\bf k}) 
{\bf U}^{\dagger} ({\bf k})
$
with matrix elements 
$
{\bf H}_{D, \alpha \beta} ({\bf k}) 
=
{\mathcal E}_{\alpha} ({\bf k}) \delta_{\alpha \beta},
$ 
where $\mathcal{E}_{\alpha} ({\bf k})$ are the eigenvalues of 
${\bf H}_{0} ({\bf k})$ discussed above.
The three-component spinor in the generalized 
helicity eigenbasis is  
$
{\bf \Phi}^{\dagger} ({\bf k}) 
= 
\left[
\phi^{\dagger}_{\Uparrow} ({\bf k}),
\phi^{\dagger}_{0} ({\bf k}),
\phi^{\dagger}_{\Downarrow} ({\bf k})
\right],
$ 
where $\phi^{\dagger}_{\alpha} ({\bf k})$ is the creation 
operator of a fermion with eigenenergy $\mathcal{E}_{\alpha} ({\bf k})$
with generalized {\it spin} label $\alpha$.
The unitary matrix
\begin{eqnarray}
\label{eqn:unitary-matrix}
{\bf U}({\bf k}) 
= 
\left(
\begin{array}{c c c}
u_{\Uparrow 1} ({\bf k}) & u_{\Uparrow 2} ({\bf k}) & u_{\Uparrow 3} ({\bf k}) \\
u_{0 1} ({\bf k}) & u_{0 2} ({\bf k}) & u_{0 3} ({\bf k}) \\
u_{\Downarrow 1} ({\bf k}) & u_{\Downarrow 2} ({\bf k}) & u_{\Downarrow 3} ({\bf k})  
\end{array}
\right)
\end{eqnarray}
has rows that satisfy the normalization condition
$
\sum_{s}
\vert u_{\alpha s} ({\bf k}) \vert^2 
= 
1,
$
where $\alpha = \{\Uparrow, 0, \Downarrow \}$.

We use as units 
the Fermi energy 
$
E_F = k_F^2/(2m)
$
and the Fermi momentum 
$
k_F = (2 \pi^2 n)^{1/3},
$
based on the total density 
of fermions $n = 3 k_F^3/(6 \pi^2)$ with initial identical
kinetic energies $\epsilon_{\bf k} = {\bf k}^2/(2m)$ 
for all three internal states. 
This means that our reference system is that with all parameters
$\eta, k_T, \Omega$ and $\delta$ set to zero. 

\begin{figure} [hbt]
\includegraphics[width = \linewidth]{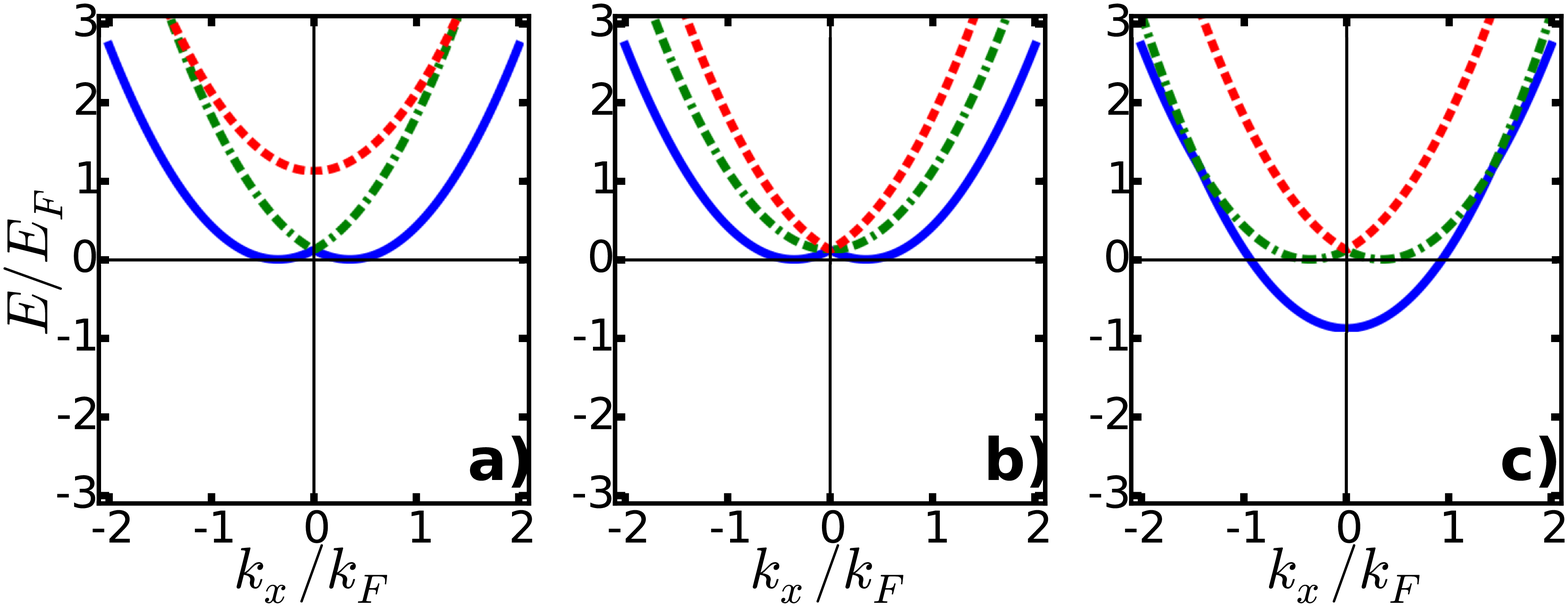}
\includegraphics[width = \linewidth]{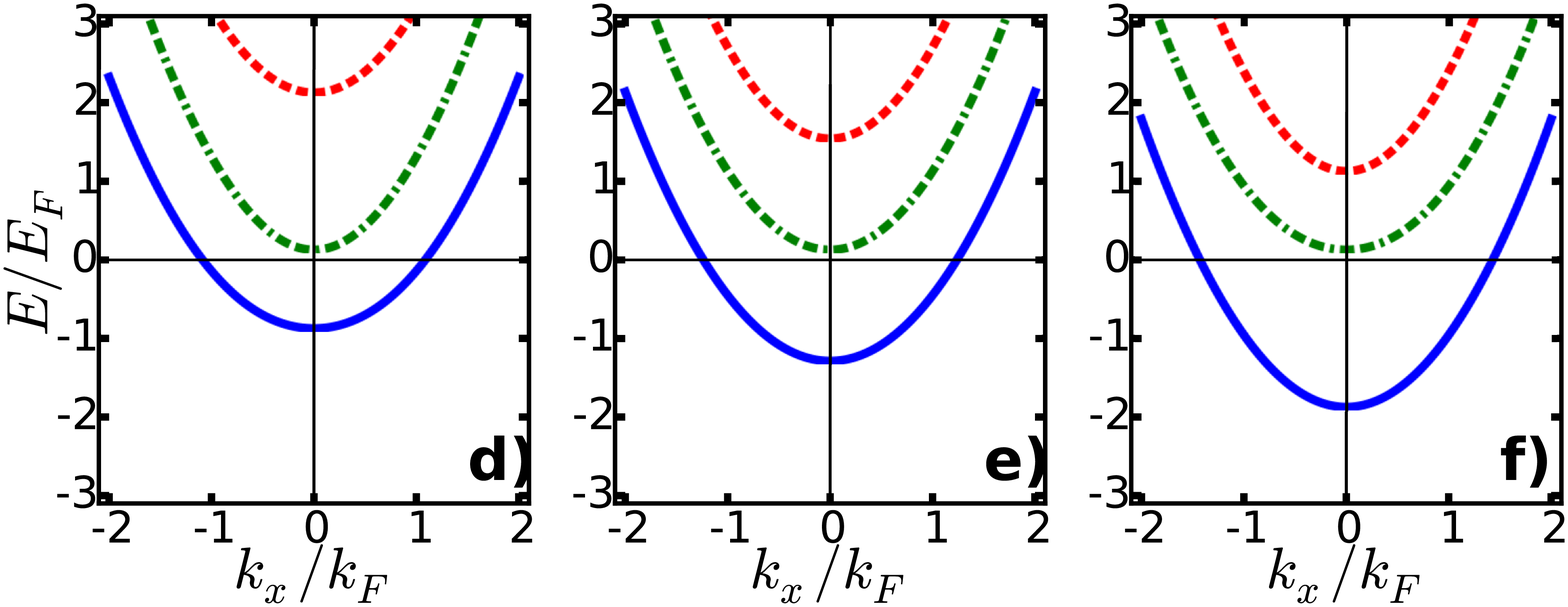}
\caption{
\label{fig:one}  
(color online) 
Eigenvalues $\mathcal{E}_{\alpha} ({\bf k})$ versus $k_x$ 
in qualitatively different situations 
corresponding to momentum transfer $k_T = 0.35k_F$, 
and quadratic Zeeman shift 
$b_z =  -E_F$ (left);
$b_z = 0$ (middle);
$b_z = E_F$ (right). 
The top (bottom) panels correspond to 
Rabi frequency $\Omega = 0$ $(\Omega = E_F)$. 
The solid blue line corresponds to $\mathcal{E}_{\Uparrow} ({\bf k})$,
the dot-dashed-green line to $\mathcal {E}_0 ({\bf k})$
and the dashed-red line to $\mathcal{E}_{\Downarrow} ({\bf k})$,
aslo at $T = 0.02 E_F$.}
\end{figure}

In Fig.~{\ref{fig:one}}, we show plots of 
eigenvalues $\mathcal{E}_{\alpha} ({\bf k})$ versus 
momentum $k_x$ for fixed momentum transfer $k_T = 0.35 k_F$
and zero detuning $\delta = 0$. The figures at left, middle, right
represent, respectively, 
the cases of quadratic Zeeman shift $b_z = -E_F, 0, +E_F$.
The top panels correspond to $\Omega = 0$, and the bottom panels to
$\Omega = E_F$. In the top panels, notice that double minima 
are present in $\mathcal{E}_{\Uparrow} ({\bf k})$ 
when $b_z < 0$ and $b_z = 0$ and that double minima appear 
in $\mathcal{E}_0 ({\bf k})$, when $b_z > 0$. This occurs 
only for small $\Omega/E_F$. In the bottom 
panels, notice that double minima in all dispersions disapappear, 
because the Rabi frequency $\Omega = E_F$ is sufficiently large.

%

In order to study the quantum phases of three-component fermions, we add
interactions between atoms in different internal states and consider
contact attractive interactions 
$-g_{ss^{\prime}} \delta ({\bf r} - {\bf r}^{\prime})$ of 
strength $g_{s s^{\prime}} > 0$, between internal states $s \ne s^{\prime}$
only. Notice that $g_{s s^{\prime}}$ has dimensions of energy times volume. 
The atom-atom interaction Hamiltonian can be written in momentum space
as
\begin{eqnarray}
\label{eqn:atom-atom-interaction-hamiltonian}
H_{{\rm AA}} 
= 
-
\frac{1}{V} 
\sum \limits_{{\bf Q},\left\{s \neq s^{\prime}\right\}} 
g_{ss^{\prime}} 
b_{ss^{\prime}}^{\dagger}({\bf Q})
b_{ss^{\prime}}({\bf Q}),
\end{eqnarray}
where 
$V$ is the volume, ${\bf Q}$ 
is the center-of-mass momentum of fermion pairs characterized
by the operator 
$
b_{ss^{\prime}}^{\dagger}({\bf Q}) 
= 
\sum_{\bf k} 
\psi_s^{\dagger}({\bf k} + {\bf Q}/2)
\psi_{s^{\prime}}^{\dagger}(-{\bf k} - {\bf Q}/2).
$ 
Thus, the Hamiltonian describing the effects of spin-orbit coupling,
Zeeman fields and atom-atom interactions is 
\begin{equation}
H
=
H_{{\rm CA}}
+
H_{{\rm AA}}
-
\mu {\hat N},
\end{equation}
where 
$
{\hat N} 
= 
\sum_{s,{\bf k}}
\psi_s^{\dagger}({\bf k})
\psi_s({\bf k})
$ 
represents the total number of particles. 
We are interested in uniform superfluid phases, so we focus on
the case of pairing at zero center-of-mass momentum only, that
is, ${\bf Q} = 0$. 

We define the order parameter for superfluidity as the 
expectation value
$
\Delta_{ss^{\prime}}
= 
-g_{ss^{\prime}}
\langle 
b_{ss^{\prime}}
(
{\bf 0}
)
\rangle/V,
$
and write the reduced mean-field Hamiltonian as
\begin{equation}
\label{eqn:mean-field-Hamiltonian}
H_{\rm MF} 
= 
\frac{1}{2}
\sum_{\bf k}
{\bf \Psi}^{\dagger}_{N} ({\bf k})
{\bf H}_{\rm MF}({\bf k})
{\bf \Psi}_{N} ({\bf k})
+
V \sum_{s\neq s^{\prime}}
\frac{|\Delta_{ss^{\prime}}|^2}{g_{ss^{\prime}}}
+
{\mathcal C} (\mu) 
\end{equation}
where the six-component Nambu spinor is 
$
{\bf \Psi}^{\dagger}_{N} ({\bf k})
=
\left[
\Psi_1^{\dagger}({\bf k}),
\Psi_2^{\dagger}({\bf k}),
\Psi_3^{\dagger}({\bf k}),
\Psi_1 ({-\bf k}),
\Psi_2 ({-\bf k}),
\Psi_3 ({-\bf k})
\right]
$
and the function
$
{\mathcal C} (\mu) 
= 
\frac{1}{2}
\sum_{{\bf k} s}
\xi_s(-{\bf k})
$
contains the term 
$
\xi_s({\bf k}) 
= 
\varepsilon_s({\bf k}) -\mu
$
representing the residual kinetic energy contributions. 
Here, 
$
\varepsilon_s({\bf k})
$
represents the diagonal matrix elements of ${\bf H}_0 ({\bf k})$ 
given by
$
\varepsilon_1 ({\bf k}) 
= 
\varepsilon ({\bf k})
- 
h_z ({\bf k})
+
b_z;
$
$
\varepsilon_2 ({\bf k}) 
= 
\varepsilon ({\bf k});
$
and
$
\varepsilon_3 ({\bf k}) 
= 
\varepsilon ({\bf k})
+
h_z ({\bf k})
+
b_z.
$

The 
$
6 \times 6
$
mean-field Hamiltonian matrix is
\begin{eqnarray}
\label{eqn:mean-field-hamiltonian-matrix}
{\bf H}_{\rm MF}({\bf k})
=
\left(
\begin{array}{cc}
\overline {\bf H}_0({\bf k})	&	{\bf \Lambda}		\\
{\bf \Lambda}^{\dagger}	&	-\overline {\bf H}_0^* (-{\bf k})
\end{array}
\right),
\end{eqnarray}
where the diagonal block matrix 
$
\overline {\bf H}_0({\bf k})
=
{\bf H}_0({\bf k})
-\mu {\bf 1}
$
and the off-diagonal block matrix 
\begin{eqnarray}
\label{eqn:order-parameter-tensor}
{\bf \Lambda}
=
\left(
\begin{array}{ccc}
0		&	\Delta_{12}	&	\Delta_{13}	\\
-\Delta_{12}	&	0		&	\Delta_{23}	\\
-\Delta_{13}	&	-\Delta_{23}	& 	0
\end{array}
\right)
\end{eqnarray}
represents the order parameter tensor. 
In this work, we consider the simpler case 
where $g_{12} = g_{23} = 0$, and $g_{13} = g$, 
which leads to $\Delta_{12} = \Delta_{23} = 0$,
and $\Delta_{13} = \Delta$, such that the order parameter tensor
$\Delta_{ss^\prime}$ is characterized by a single complex scalar $\Delta$.

The quasi-particle and quasi-hole excitation spectrum can be 
found from the mean-field Hamiltonian 
in Eq.~(\ref{eqn:mean-field-hamiltonian-matrix})
leading to six energy eigenvalues, which we order as  
$
E_1({\bf k})
>
E_2({\bf k})
>
E_3({\bf k})
>
E_4({\bf k})
>
E_5({\bf k})
>
E_6({\bf k}).
$
These eigenvalues exhibit quasi-particle/quasi-hole symmetry
in momentum space for any value of detuning $\delta$ and Rabi frequency
$\Omega$, which means 
$
E_6({\bf k}) = -E_1(-{\bf k}),
$
$
E_5({\bf k}) = -E_2(-{\bf k})
$
and
$
E_4({\bf k}) = -E_3(-{\bf k}).
$
However, each eigenergy $E_j ({\bf k})$ has well defined parity only
when $\delta = 0$, in which case $E_j ({\bf k}) = E_j ({\bf - k})$ 
has even parity. 

To analyze the excitation spectrum $E_j ({\bf k})$, we need to determine
self-consistently the values of the order parameter amplitude 
$\Delta_{13} = \Delta$ and the chemical potential $\mu$. 
For this purpose, we write the thermodynamic potential
$
\mathcal{Q}
= 
-T
\ln
\mathcal{Z},
$
where 
$
\mathcal{Z} 
= 
\int
\Pi_{s}D
\left[
\psi^{\dagger}_s ({\bf k})
,
\psi_s ({\bf k})
\right]
\exp
\left[
-S
\right]
$
is the grand-canonical partition function written in terms of the
action $S$. At the mean-field level the action is
$$
T S_{\rm MF} 
= 
-
\frac{1}{2}
\sum_{n,{\bf k} }
{\bf \Psi}_{N}^{\dagger} ({\bf k})
{\bf G}^{-1} 
{\bf \Psi}_{N} ({\bf k})
+
V \frac{\vert \Delta_{13} \vert^2}{g_{13}}
+
{\mathcal C} (\mu), 
$$
where 
$
{\bf G}^{-1} (i\omega_n, {\bf k})
=
\left[
i\omega_n {\bf 1}
-
{\bf H}_{\rm MF}({\bf k})
\right]
$
is the inverse of the resolvent (Green) matrix 
${\bf G} (i\omega_n, {\bf k})$. Here, 
$
\omega_n = (2n+1)\pi T
$
is the fermionic Matsubara frequency, and 
$T$ is the temperature. 
Integration over the fermionic fields leads to the
thermodynamic potential 
\begin{equation}
{\mathcal Q}_{\rm MF}
=
-\frac{T}{2}
\sum_{{\bf k} j} 
\ln 
\left\{
1 
+
\exp \left[ - E_j ({\bf k})/T \right] 
\right\}
+
V \frac{\vert \Delta_{13} \vert^2}{g_{13}}
+
{\mathcal C} (\mu), 
\end{equation}
where the sum over $j$ includes both quasi-particle and quasi-hole energies
$(j = \{1, 2, 3, 4, 5, 6 \})$.

The minimization of  
${\mathcal Q}_{\rm MF}$ with respect to $\Delta_{13}^*$ using
the relation $\delta {\mathcal Q}_{\rm MF}/\delta \Delta_{13}^* = 0$
leads to the order parameter equation
\begin{equation}
\label{eqn:order-parameter-equation}
\frac{V}{g_{13}} \Delta_{13}
=
\frac{1}{2} 
\sum \limits_{\bf k}
\sum \limits_{j=1}^3
\tanh\left(
\frac{\beta E_j({\bf k})}{2}
\right) 
\frac{\partial E_j({\bf k})}{\partial \Delta_{13}^*},
\end{equation}
while fixing the total number of particles for $\Omega \ne 0$ is possible
via the thermodynamic relation 
$N = - \partial{\mathcal Q}_{\rm MF}/\partial \mu \vert_{T,V}$
leading to the number equation
\begin{equation}
\label{eqn:number-equation}
N 
=  
\frac{1}{2}
\sum \limits_{\bf k}
\left[
\sum \limits_{j=1}^3
\tanh\left(
\frac{\beta E_j({\bf k})}{2}
\right) 
\frac{\partial E_j({\bf k})}{\partial \mu}
+
3
\right].
\end{equation}
Here, the summation over $j$ involves only 
quasi-particle energies $(j = \{1, 2, 3\})$, 
that is, only the positive eigenvalues of the Hamiltonian matrix
${\bf H}_{\rm MF}({\bf k})$, because we used quasi-particle/quasi-hole 
symmetry to eliminate the quasi-hole energies. 
Lastly, by using the relation
$
V/g_{13} 
= 
-mV/(4 \pi a_s) 
+ 
\sum_{\bf k}1/(2 \epsilon_{\bf k}),
$
we express the 
bare coupling constant $g_{13}$ in terms of the scattering length $a_s$ 
in the absence of the spin-orbit and Zeeman fields.

Among the quasi-particle excitation energies 
$ 
E_1({\bf k}), 
$
$
E_2({\bf k}),
$
and 
$
E_3({\bf k}),
$
only $E_3({\bf k})$ may have zeros. 
In momentum space, the zeros of $E_3 ({\bf k})$ define the 
{\it loci} (points, lines or surfaces) 
where there is no energy cost to create quasi-particle excitations. 
Such points, lines or surfaces
of zero energy can be used to classify the topologically distinct 
superfluid phases of three-component fermions.

\begin{figure} [hbt]
\centering
\epsfig{file=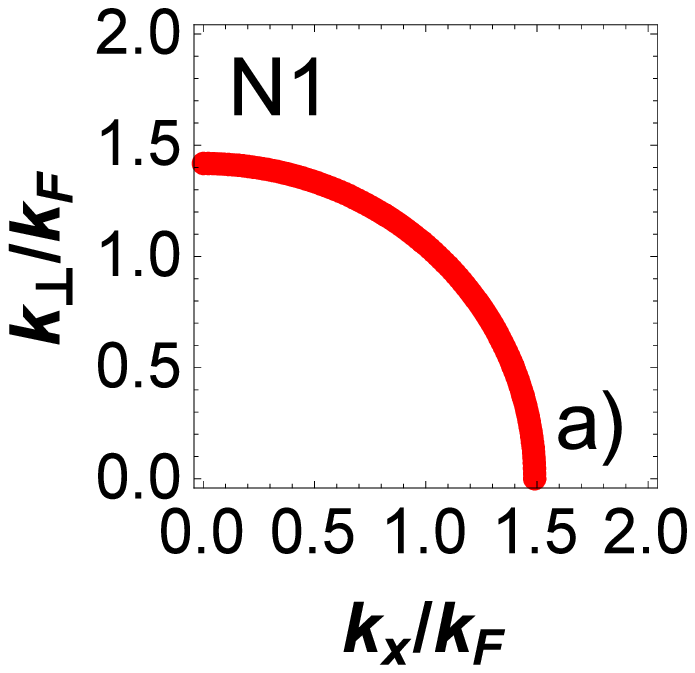,width=0.32 \linewidth}
\epsfig{file=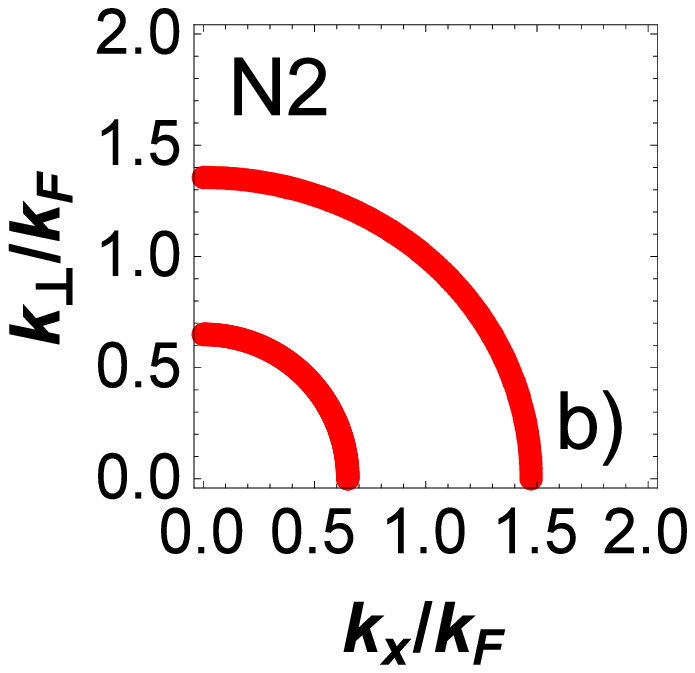,width=0.32 \linewidth}
\epsfig{file=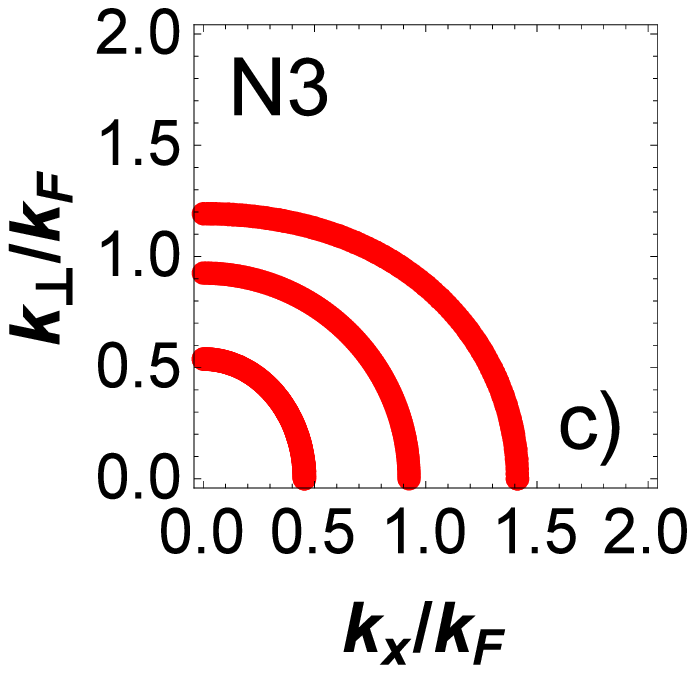,width=0.32 \linewidth}
\\
\epsfig{file=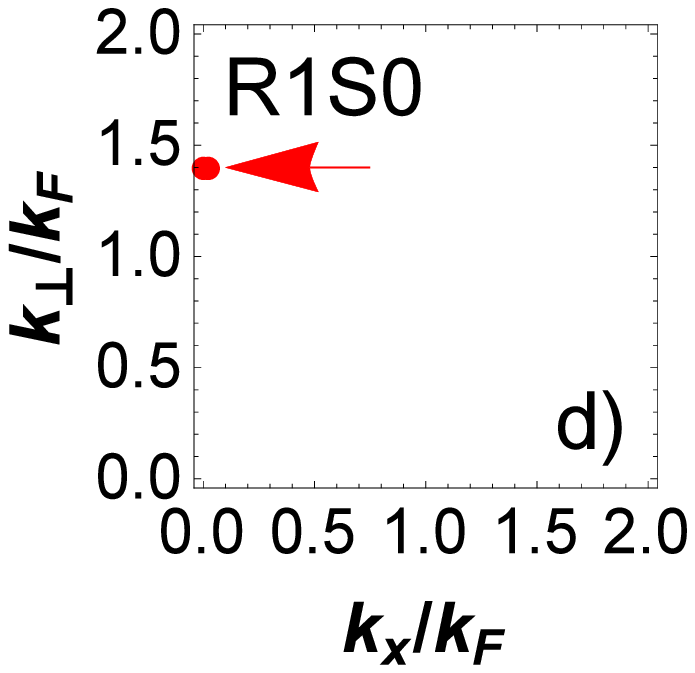,width=0.32 \linewidth}
\epsfig{file=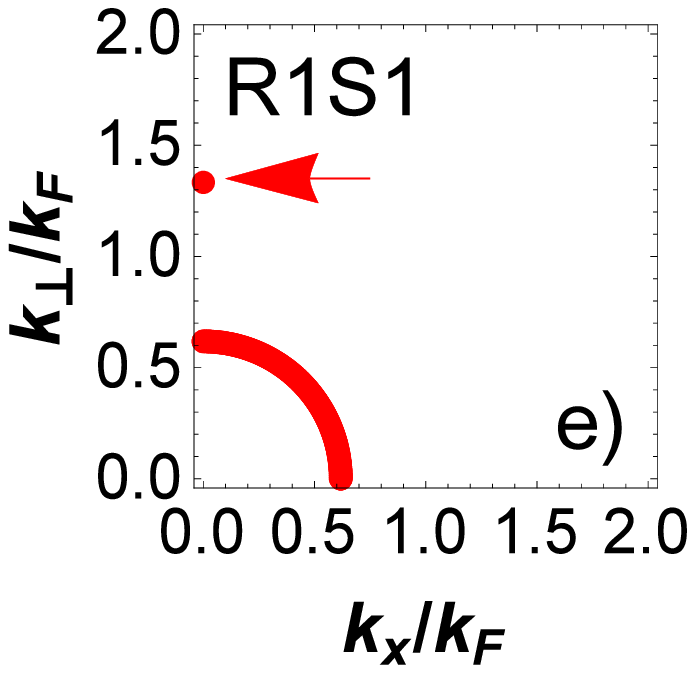,width=0.32 \linewidth}
\epsfig{file=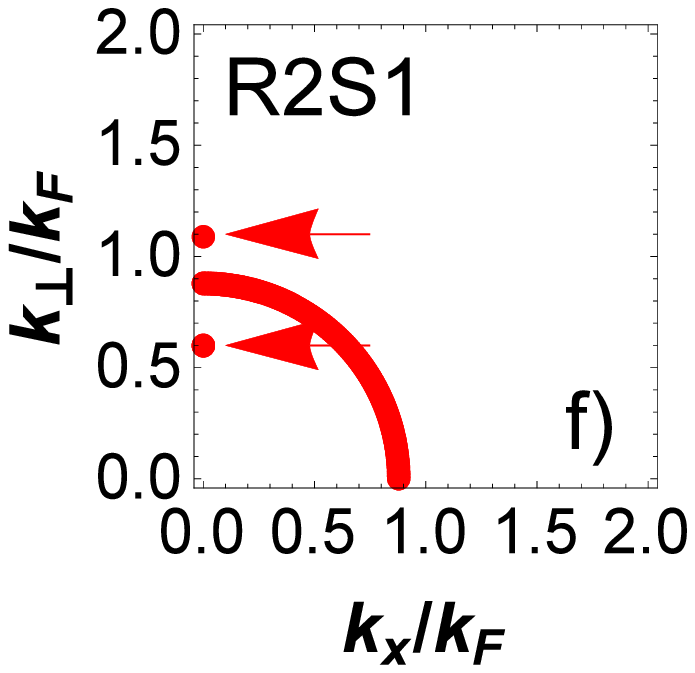,width=0.32 \linewidth}
\caption{ 
\label{fig:two}
Plots of {\it loci} of zero quasi-particle energy  
$E_3(k_x,k_{\perp})$ in the $k_xk_\perp$ plane 
at $T = 0.02 E_F$ for various phases with $b_z = 0$, 
where $k_{\perp}$ is the radial
component representing $(k_y, k_z)$. 
The phases are: 
a) $N1$ with $\Omega = 1.75E_F$ and $1/(k_F a_s) = -1.00$,
b) $N2$ with $\Omega = E_F$ and $1/(k_F a_s) = -1.00$,
c) $N3$ with $\Omega = 0.40E_F$ and $1/(k_F a_s) = -1.00$,
d) $R1S0$ with $\Omega = 1.75E_F$ and $1/(k_F a_s) = 0.80$,
e) $R1S1$ with $\Omega = E_F$ and $1/(k_F a_s) = 0.20$,
f) $R2S1$ with $\Omega = 0.40E_F$ and $1/(k_F a_s) = -0.30$.
}
\end{figure}

In Fig.~\ref{fig:two},
we plot the momentum space {\it loci} of
$E_3 ({\bf k}) = 0$ versus $(k_x, k_{\perp})$, where $k_{\perp}$
represents a radial vector in the $k_y k_z$ plane,
for the case of zero detuning $(\delta = 0)$.
We show only the first quadrant, because the {\it loci} have 
polar symmetry in the  $k_y k_z$ plane, and 
reflection symmetry in the $k_x$ direction. 
This means that points along the $k_\perp$ axis represent circles in the
$k_y k_z$ plane, and that lines in the $k_x k_{\perp}$ represent surfaces
in three-dimensional momentum space $(k_x, k_y, k_z)$.
The top panels show the momentum space {\it loci} for normal phases 
$N1, N2, N3$, represent one, two or three distinct surfaces, respectively.
These are the Fermi surfaces associated with the normal phases.
In the bottom panels, we show the nodal structure of three superfluid phases 
that have a boundary with the normal state, when the quadratic Zeeman 
term is zero $(b_z = 0)$. The phase $R1S0$ has one ring and zero
surface of nodes, the phase $R1S1$ has one ring and one surface of nodes, 
and the phase $R2S1$ has two rings and one surface of nodes. 
For $b_z = 0$, there is also a phase with zero rings and one 
surface of nodes $(R0S1)$, not shown in Fig.~\ref{fig:two}.

%

\begin{figure} [hbt]
\centering
\epsfig{file=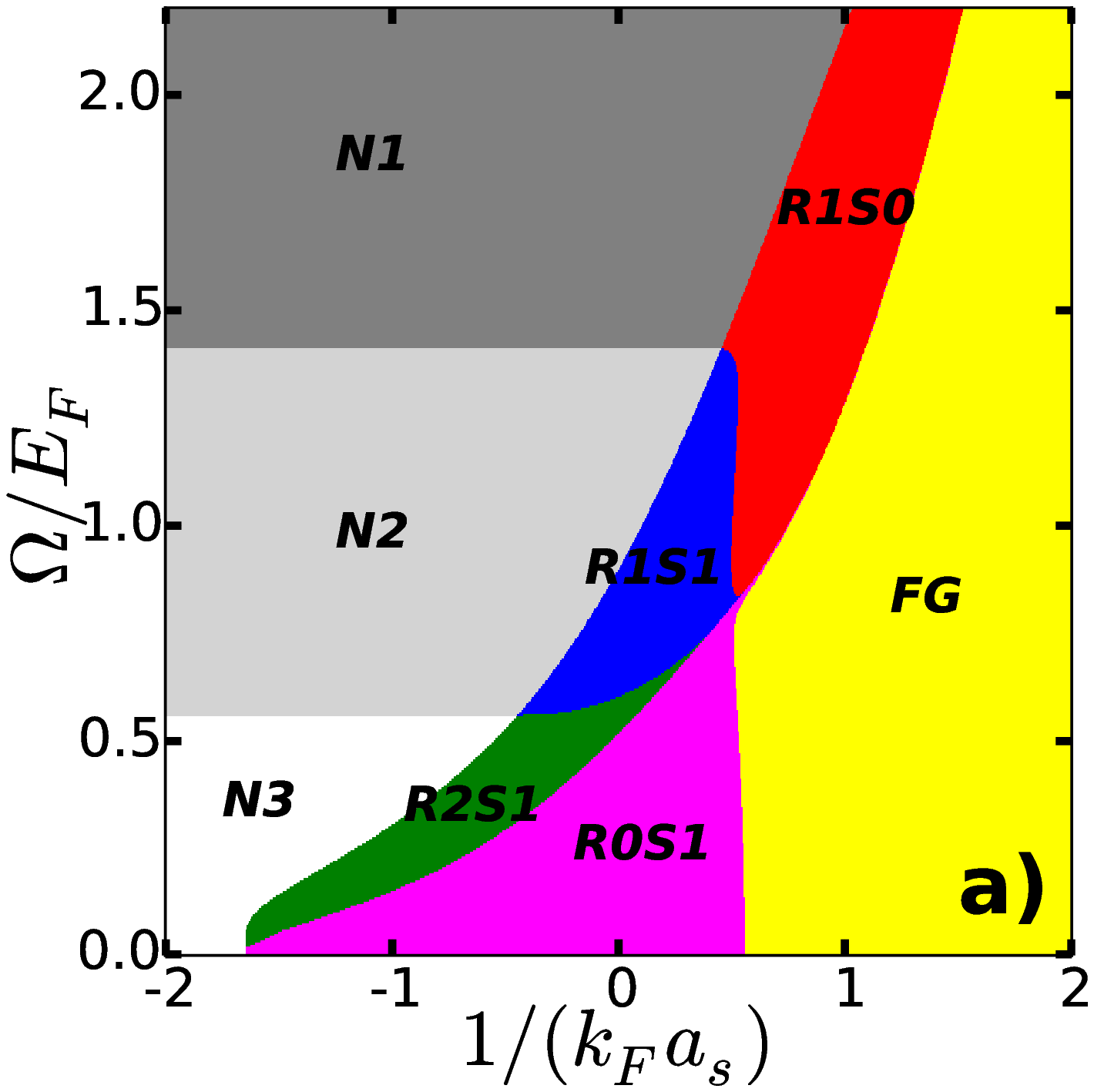,width=0.49 \linewidth}
\epsfig{file=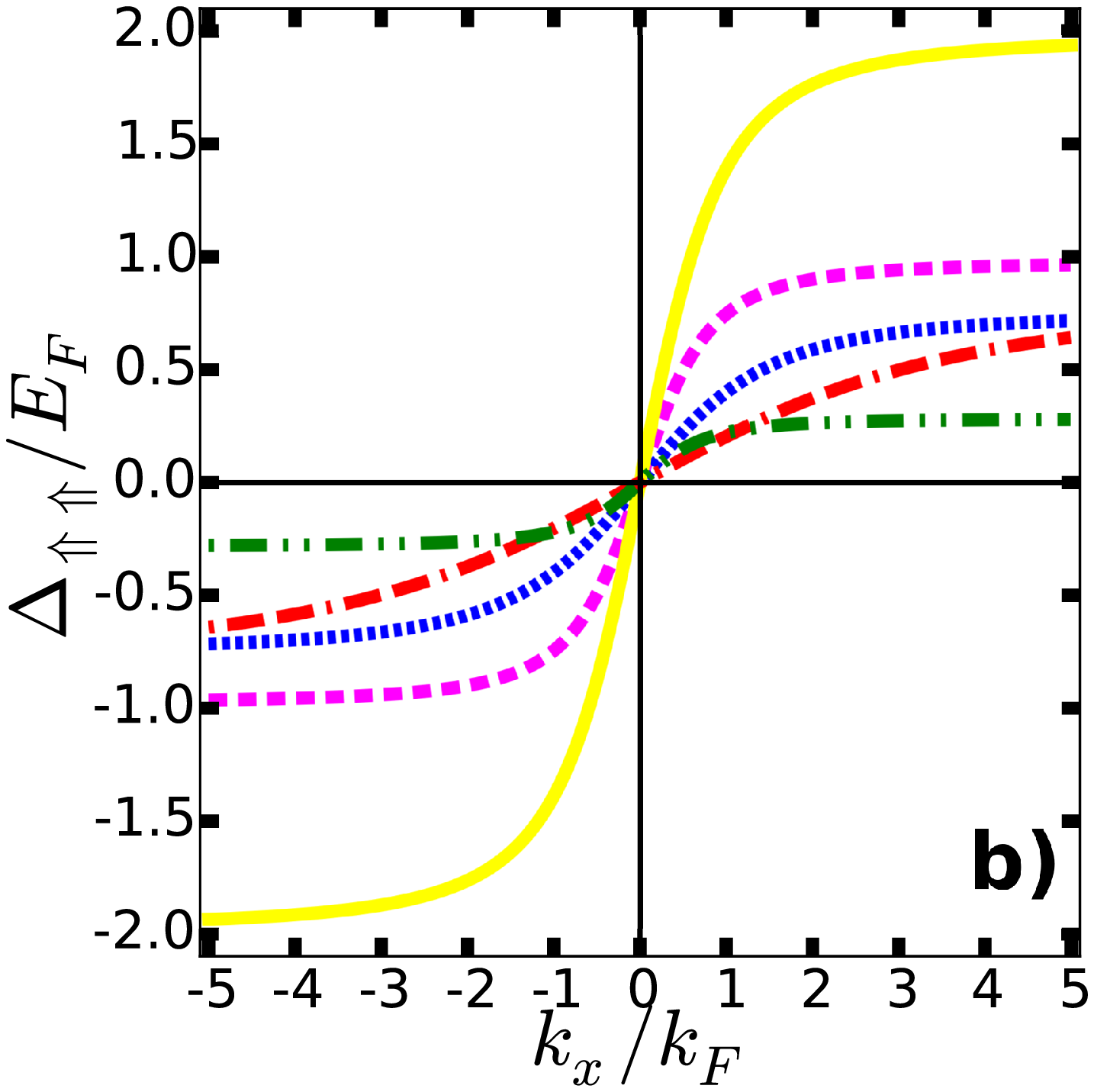,width=0.49 \linewidth}
\\
\epsfig{file=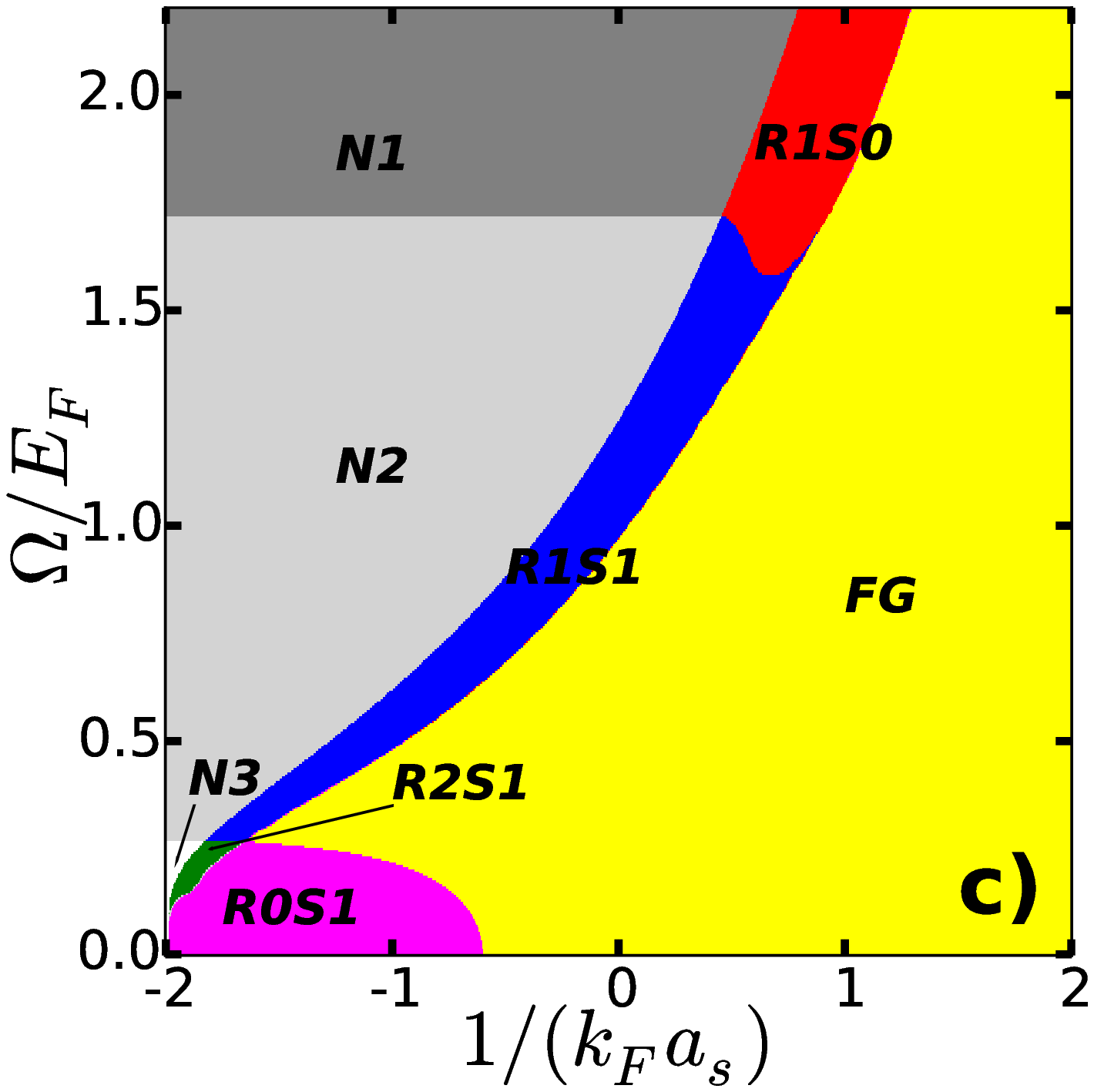,width=0.49 \linewidth}
\epsfig{file=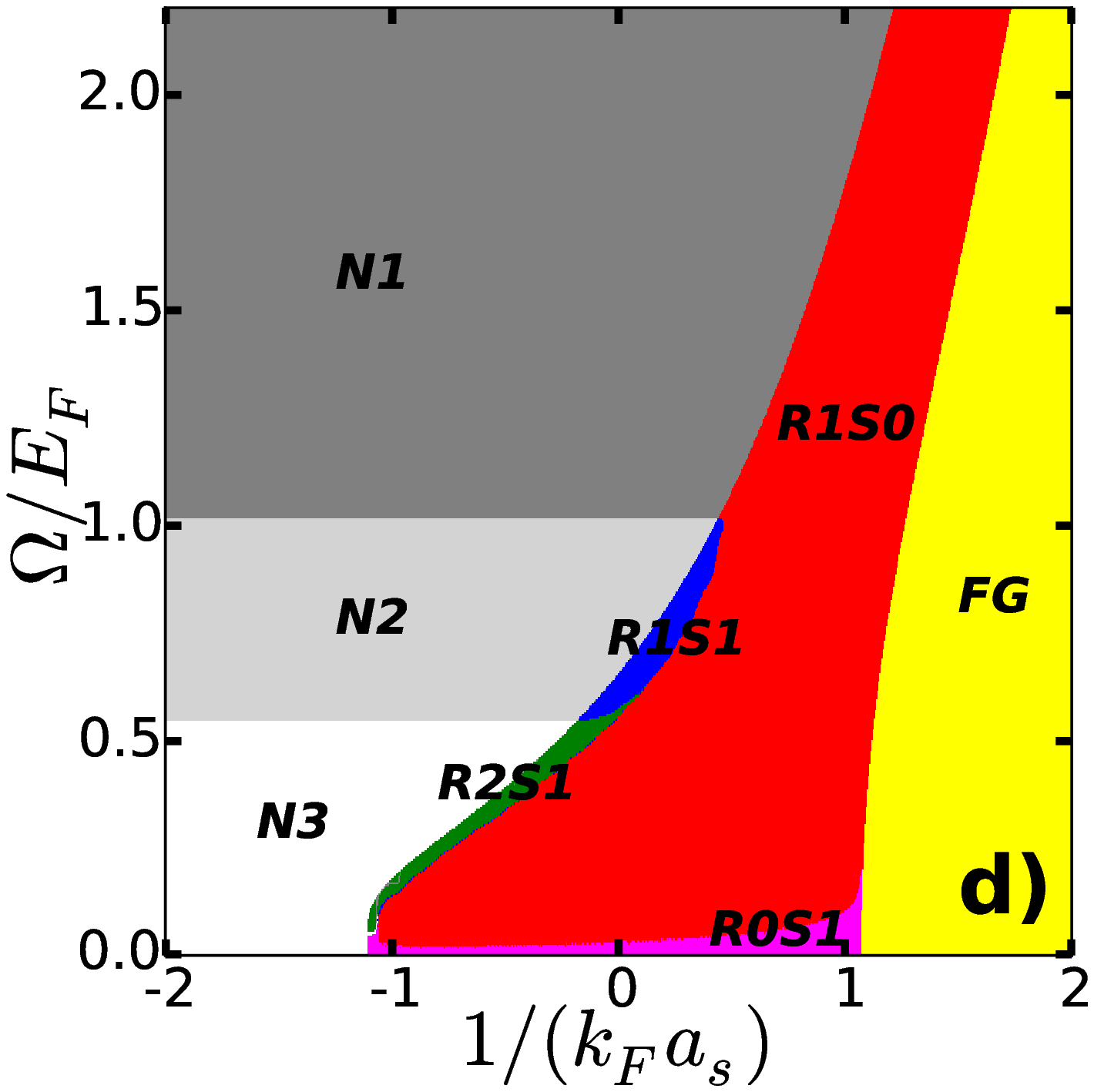,width=0.49 \linewidth}
\caption{ 
\label{fig:three}
(Color online) 
Phase diagrams of $\Omega/E_F$ versus  $1/(k_F a_s)$  
for $T = 0.02 E_F$ and $k_T = 0.35 k_F$ are shown in a) $b_z = 0$,  
c) $b_z = -E_F$, and  
d) $b_z = +E_F$. 
The normal phases $N1$, $N2$ and $N3$ 
are indicated by dark-gray, light-gray and white colors, respectively. 
The superfluid phases are color-coded as follows: $R1S0$ (red), $R1S1$ (blue), 
$R2S1$ (green), $R0S1$ (purple), and $FG$ (yellow).  
In b) we show $\Delta_{\Uparrow \Uparrow} ({\bf k})$ for
phases:
$R1S0$ (dot-double-dashed red) 
with $\Omega/E_F = 1.92$ and $1/(k_F a_s) = 1.00$, 
$R1S1$ (dotted blue) 
with $\Omega/E_F = 0.79$ and $1/(k_F a_s) = 0.23$, 
$R2S1$ (dashed-double-dotted green) 
with $\Omega/E_F = 0.40$ and $1/(k_F a_s) = -0.50$, 
$R0S1$ (dashed purple) 
with $\Omega/E_F = 0.42$ and $1/(k_F a_s) = 0.24$, 
$FG$ (solid yellow) 
with $\Omega/E_F = 0.50$ and $1/(k_F a_s) = 1.50$.
}
\end{figure}

In Fig.{~\ref{fig:three}}, we plot phase diagrams 
of Rabi frequency $\Omega/E_F$ versus scattering parameter $1/(k_F a_s)$
for quadratic Zeeman shifts $b_z = \{0, \pm E_F \}$
with spin-orbit coupling parameter $k_T = 0.35 k_F$. 
In Fig.{~\ref{fig:three}}b, for $b_z = 0$, 
we show the component $\Delta_{\Uparrow \Uparrow} ({\bf k})$ 
of the order parameter tensor in the generalized helicity basis to reveal 
its odd symmetry and its amplitude for large momentum $k_x$.   
Normal phases $N1$, $N2$ and $N3$ 
are indicated by dark-gray, light-gray and white colors, respectively.
Superfluid phases are color coded: $R1S0$ (red), $R1S1$ (blue), 
$R2S1$ (green), $R0S1$ (purple), and the fully gapped $(FG)$ phase (yellow).  
The phase transitions from normal to superfluid 
are all continuous (second-order), and  
the transitions between superfluid phases are topological and of the 
Lifshitz-type~\cite{lifshitz-1960}.
For $b_z = 0$ (Fig.{~\ref{fig:three}}a), 
there is a quintuple and pentacritical point, where 
the phases $R1S0$, $R1S1$, $R1S2$, and $R0S1$ 
converge into a fully gapped phase when rings and surfaces of nodes 
in momentum space disappear through zero momentum $({\bf k} = 0)$.
When $\Omega \to 0$, the number of particles in each band 
is conserved separatelly and spin-orbit coupling can be gauged away, 
leading to an inert band 2 and to standard crossover phenomena in 
the superfluid phase for bands 1 and 3.
%
%

In order to understand the momentum space topology of the 
quasi-particle and quasi-hole
spectra, it is convenient to write 
${\bf H}_{\rm MF} ({\bf k})$ 
defined Eq.~(\ref{eqn:mean-field-hamiltonian-matrix}) 
as 
\begin{eqnarray}
\label{eqn:mean-field-hamiltonian-matrix-helicity}
\widetilde{\bf H}_{\rm MF}({\bf k})
=
\left(
\begin{array}{cc}
\widetilde {\bf H}_D ({\bf k})	&	\widetilde {\bf \Lambda}
\\
\widetilde{\bf \Lambda}^{\dagger} &	-\widetilde {\bf H}_D^* (-{\bf k})
\end{array}
\right)
\end{eqnarray}
in the generalized helicity basis.
The matrix elements of $\widetilde{\bf H}_D({\bf k})$ 
are 
$
\widetilde {\bf H}_{D,\alpha \beta} ({\bf k})
=
\widetilde{\mathcal E}_{\alpha} ({\bf k})
\delta_{\alpha \beta},
$ 
with 
$
\widetilde {\mathcal E}_{\alpha} ({\bf k})
=
{\mathcal E}_{\alpha} ({\bf k})
-
\mu
$
and the matrix elements of 
$
\widetilde {\bf \Lambda}
$
are 
$
\widetilde {\bf \Lambda}_{\alpha \beta}
=
\Delta_{\alpha \beta} ({\bf k}).
$	
The matrix 
$
\widetilde {\bf \Lambda}
$
describing the order parameter tensor
$
\Delta_{\alpha \beta} ({\bf k})
$
in the generalized helicity basis
is momentum {\it dependent} in contrast
to the original matrix
$
{\bf \Lambda},
$
defined in Eq.~(\ref{eqn:order-parameter-tensor}),
which is {\it independent} of momentum. The order parameter tensor 
becomes
\begin{equation}
\Delta_{\alpha \beta} ({\bf k})
=
\Delta 
\left[
u_{\alpha 1} ({\bf k}) u_{\beta 3} ({\bf - k})
-
u_{\alpha 3} ({\bf k}) u_{\beta 1} ({\bf - k})
\right],
\end{equation}
where $u_{\gamma i} ({\bf k})$ are matrix elements
of ${\bf U}({\bf k})$ in Eq.~(\ref{eqn:unitary-matrix}), 
and represent the first $(i = 1)$ and third $(i = 3)$ 
components of the eigenvector amplitudes
${\bf u}_{\alpha} ({\bf k}) 
= 
\left[
u_{\alpha 1} ({\bf k}), u_{\alpha 2} ({\bf k}), u_{\alpha 3} ({\bf k})
\right].
$
The general property 
$
\Delta_{\alpha \beta} ({\bf k}) 
= 
- \Delta_{\beta \alpha} ({\bf - k}),
$
guarantees that the diagonal elements
$\Delta_{\alpha \alpha} ({\bf k})$ 
have odd parity, as required by the Pauli exclusion principle. 
However, this is not sufficient
to force the off-diagonal elements to have well defined parity.

In the case of the Hamiltonian matrix 
of Eq.~(\ref{eqn:spin-one-hamiltonian}) when $\Omega \ne 0$, 
the matrix elements are
%
$
u_{\alpha 1} ({\bf k})
=
\Omega {\mathcal N}_\alpha ({\bf k})
\vert B_\alpha ({\bf k}) \vert
{\rm sgn} \left[ A_{\alpha} ({\bf k}) \right]
$
%
for the first component, and
%
$
u_{\alpha 3} ({\bf k})
=
\Omega {\mathcal N}_\alpha ({\bf k})
\vert A_\alpha ({\bf k}) \vert
{\rm sgn} \left[ B_{\alpha} ({\bf k}) \right]
$
%
for the third component.
Here, we used the definitions
$
A_{\alpha} ({\bf k}) 
=
\varepsilon_{1} ({\bf k}) - {\mathcal E}_{\alpha} ({\bf k}),
$
$
B_{\alpha} ({\bf k}) 
=
\varepsilon_{3} ({\bf k}) - {\mathcal E}_{\alpha} ({\bf k}),
$
with
$
{\mathcal N}_{\alpha} ({\bf k}) 
=
\left\{
A_{\alpha}^2 ({\bf k}) B_{\alpha}^2 ({\bf k}) 
+
\Omega^2
\left[
A_{\alpha}^2 ({\bf k}) 
+  
B_{\alpha}^2 ({\bf k}) 
\right]
\right\}^{-1/2}
$
being a normalization function. 
Neither 
$
\varepsilon_{1} ({\bf k}) 
$
nor 
$
\varepsilon_{3} ({\bf k}) 
$
have well defined parity unless both the detuning $\delta = 0$ 
and the spin-orbit coupling parameter $k_T = 0$. 
However, for $\delta = 0$, 
additional symmetries emerge in 
$\Delta_{\alpha \beta} ({\bf k})$, because the field 
$h_z ({\bf k}) = 2k_T k_x/(2m)$ has now odd parity. This 
leads to the relation
$
\varepsilon_1 ({\bf - k}) 
= 
\varepsilon_{3} ({\bf k}), 
$
which makes 
$
A_{\alpha} ({\bf - k}) 
= 
B_{\alpha} ({\bf k})
$ 
forces $u_{\alpha 1} ({\bf - k}) = u_{\alpha 3}({\bf k})$
and leads to a symmetric order parameter tensor: 
$\Delta_{\alpha \beta} ({\bf k}) = \Delta_{\beta \alpha} ({\bf k})$.

\begin{figure} [hbt]
\centering 
\epsfig{file=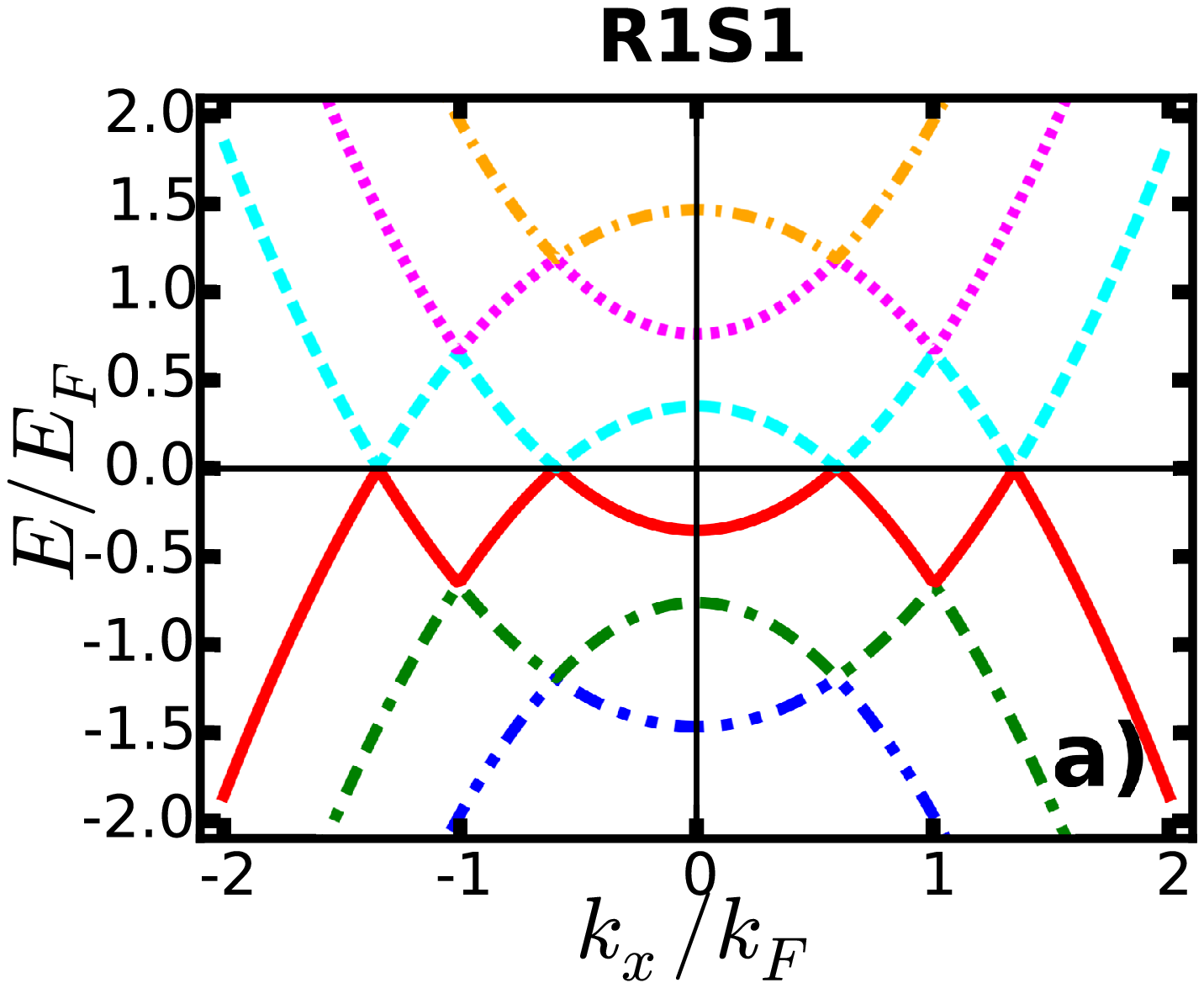,width=0.49 \linewidth}
\epsfig{file=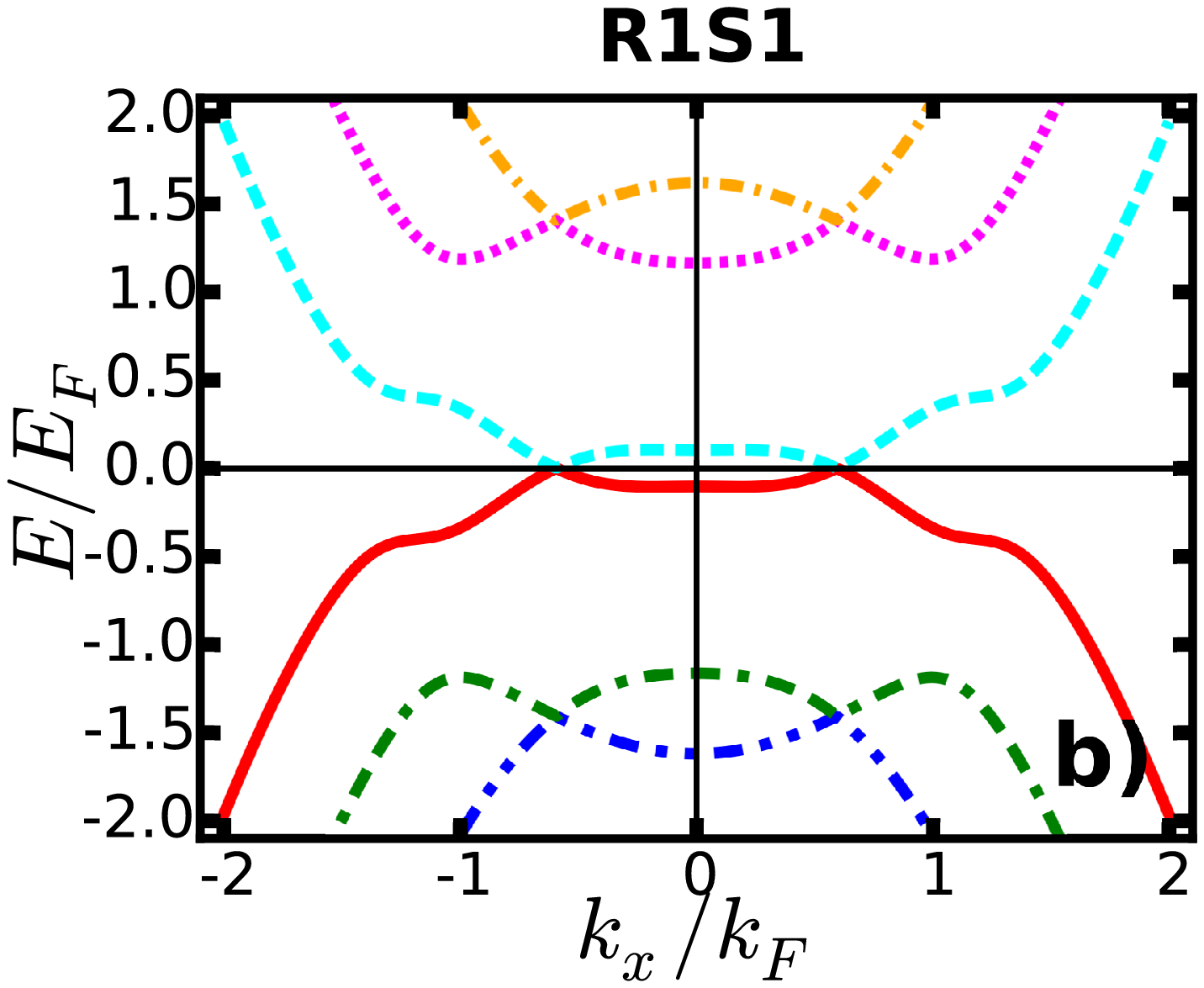,width=0.49 \linewidth}
\\
\epsfig{file=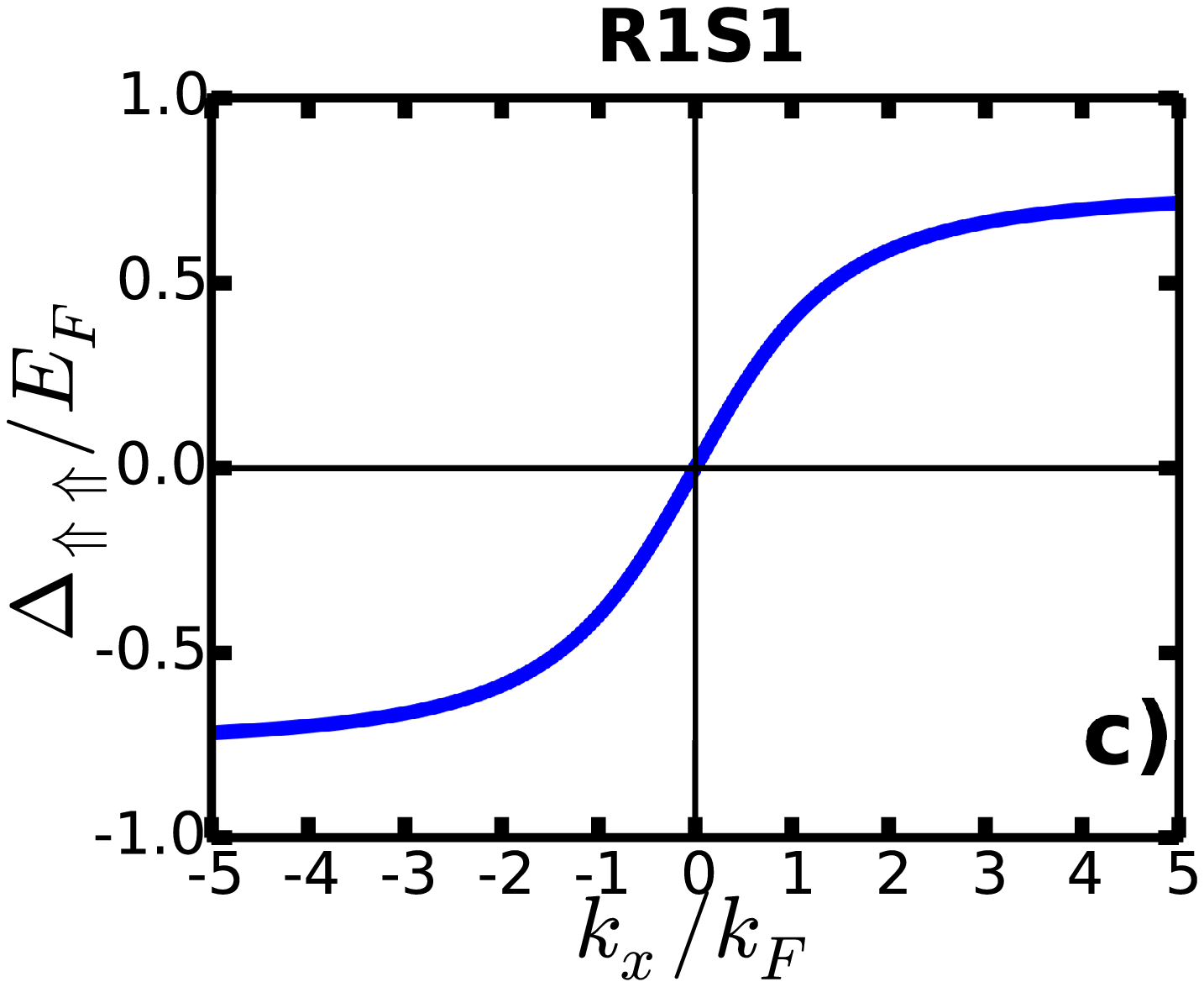,width=0.49 \linewidth}
\epsfig{file=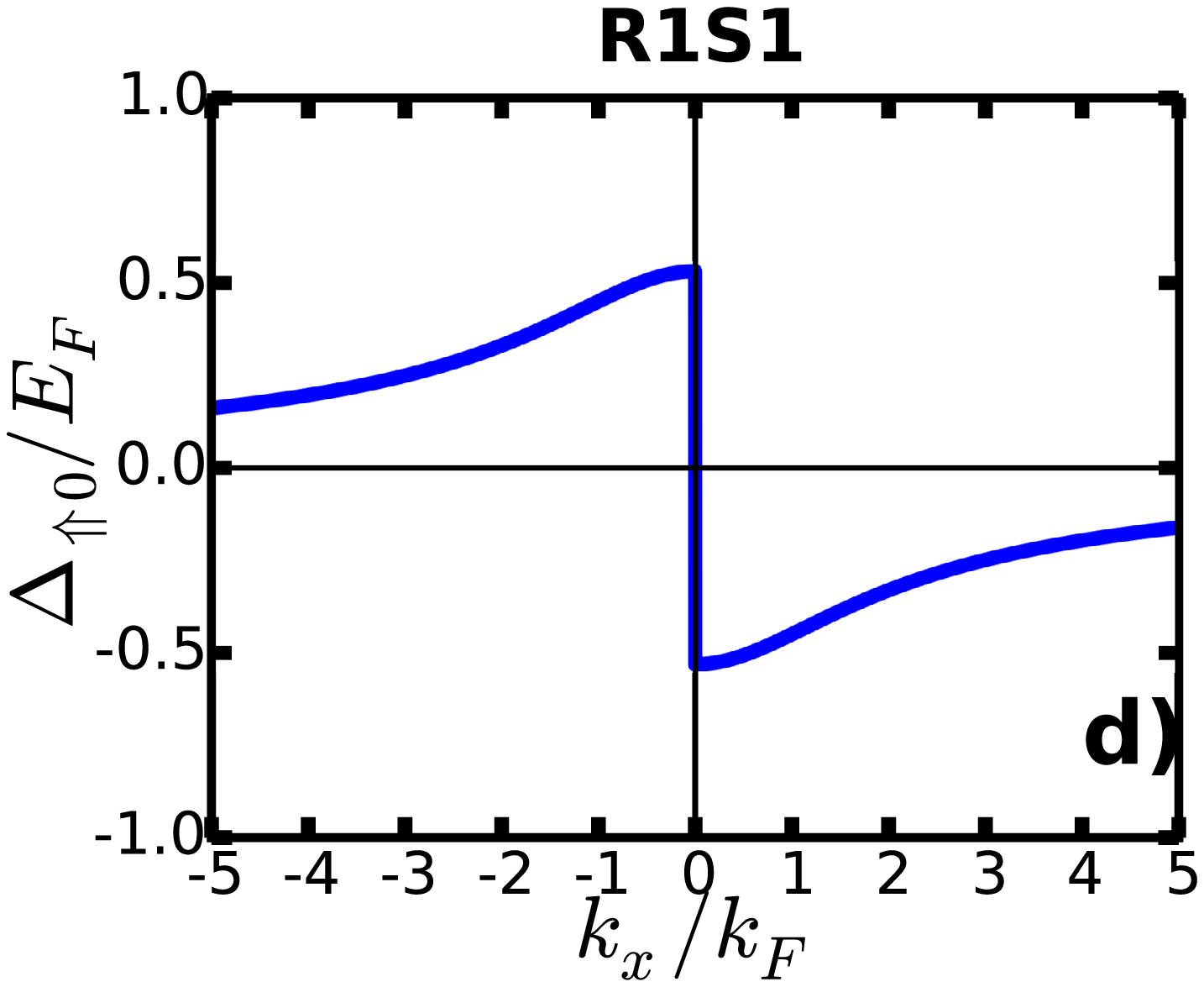,width=0.49 \linewidth}
\caption{ 
\label{fig:four}
(Color online) 
The quasi-particle and quasi-hole energies versus momentum $k_x$ 
at $T = 0.02 E_F$ are shown in a) for $\Delta = 0$ 
and in b) for $\Delta \ne 0$, corresponding to the phase $R1S1$ 
with $\Omega/E_F = 0.79$ and $1/(k_F a_s) = 0.23$.
The quasi-particle energies are shown as dot-dashed gold, dotted purple,
and dashed cyan lines. The quasi-hole energies are shown as double-dot-dashed
blue, double-dash-dotted green, and solid red.
The corresponding order parameter tensor components 
$\Delta_{\Uparrow \Uparrow} ({\bf k})$ and $\Delta_{\Uparrow 0} ({\bf k})$
versus $k_x$ are shown in c) and d).
}
\end{figure}

In Fig.~\ref{fig:four}, we show in a) the quasi-particle and quasi-hole 
excitation spectra versus momentum $k_x$ with $\Delta = 0$,
which are described by the energies 
$\widetilde{\mathcal E}_{\alpha} ({\bf k})$ 
and 
$-\widetilde{\mathcal E}_{\alpha} (-{\bf k})$,
and in b) we show the corresponding spectra for the superfluid phase $R1S1$, 
with $\Omega/E_F = 0.79$ and $1/(k_F a_s) = 0.23$. 
In c) and d) we show the order parameter tensor components 
$\Delta_{\Uparrow\Uparrow} ({\bf k})$ and $\Delta_{\Uparrow 0} ({\bf k})$ 
versus $k_x$ to illustrate how the gaps 
in the excitation spectra of b) emerge from the spectra of a)
by lifting appropriate degeneracies. 
%

We have studied the quantum phases of interacting three component fermions
in the presence of spin-orbit coupling and Zeeman fields. We classified the
emerging superfluid phases in terms of the the {\it loci} of zeros 
of their quasi-particle excitation spectrum in momentum space, 
and identified several Lifshitz-type topological transitions. In the
particular case that the quadratic Zeeman field is zero, a quintuple point
exists where five gapless superfluid phases with surface and line nodes
converge into a fully gapped superfluid phase. Lastly, we also showed that 
the simultaneous presence of spin-orbit and Zeeman fields transforms 
a momentum-independent scalar order parameter into an explicitly 
momentum-dependent tensor in the generalized helicity basis.

\acknowledgments{C. A. R. SdM acknowledges the support of the 
Joint Quantum Institute during a sabbatical visit.}

\clearpage
\newpage

\begin{figure}[htb]
\includegraphics[width = 2.00\linewidth]{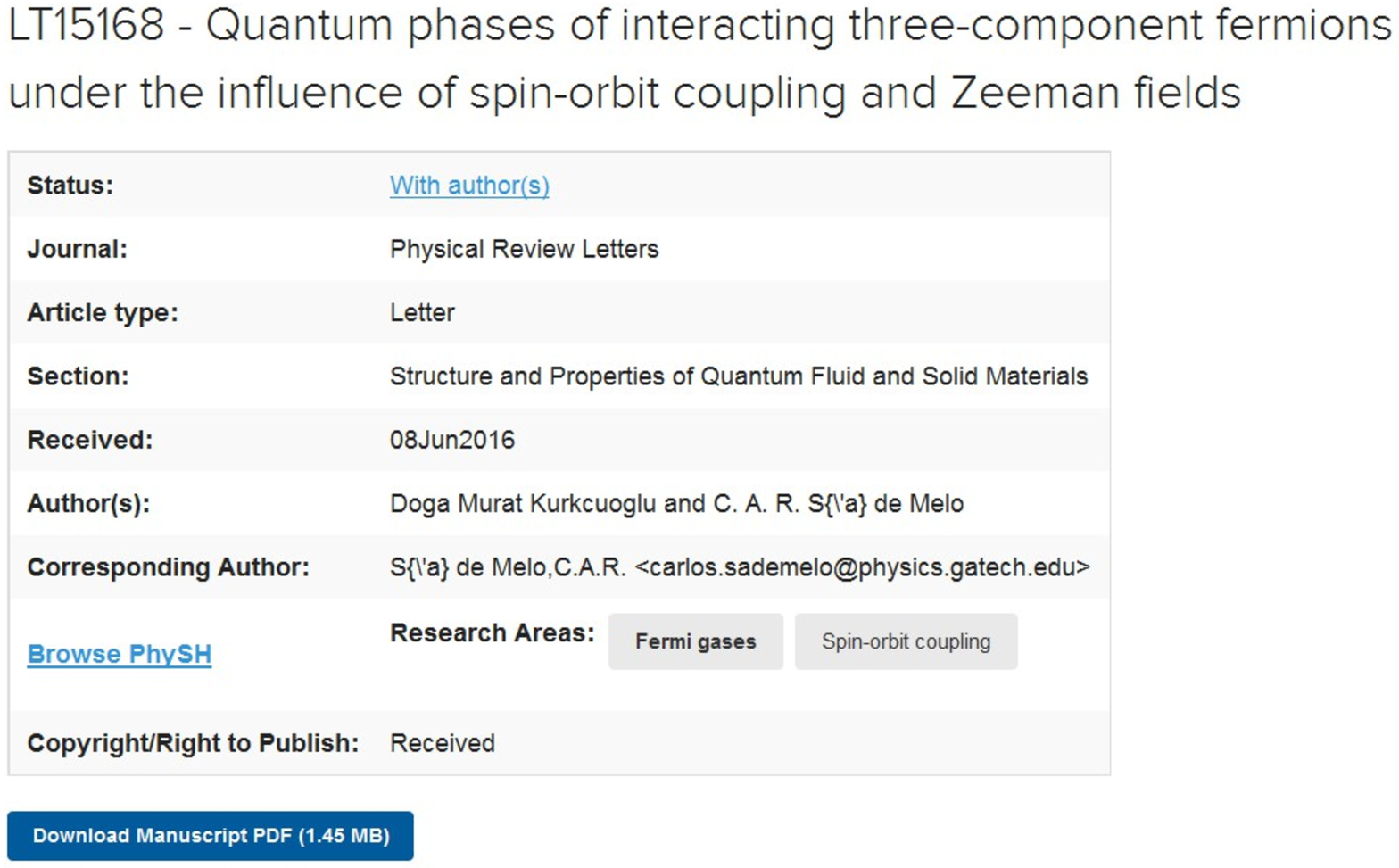}
\end{figure}

\end{document}